\documentclass[preprint,aps,prd,superscriptaddress,showpacs]{revtex4}

\usepackage{amsmath}
\usepackage{amssymb}
\usepackage{amsfonts}

\def\PBG#1#2{\{#1, #2\}^G}
\def\PBGamma#1#2{\{#1, #2\}^\Gamma}

\def\C{\mathcal{C}}

\def\d{\mathrm{d}}

\def\H{\mathcal{H}}

\def\A{\mathcal{A}}
\def\B{\mathcal{B}}

\def\nab{\nabla\!_\gamma{}}

\begin{document}


\title{Conformal decomposition in canonical general relativity}

\author{Charles H.-T. Wang}
\email{c.wang@abdn.ac.uk}
\affiliation{School of Engineering and Physical Sciences,
University of Aberdeen, King's College, Aberdeen AB24 3UE, Scotland}
\affiliation{Space Science and Technology Department,
Rutherford Appleton Laboratory, Didcot, Oxon OX11 0QX, England}


\begin{abstract}
A new canonical transformation is found that enables the direct
canonical treatment of the conformal factor in general relativity.
The resulting formulation significantly simplifies the previously
presented conformal geometrodynamics and provides a further
theoretical basis for the conformal approach to loop quantum
gravity.
\end{abstract}


\maketitle

\section{Background}

The well-known Arnowitt-Deser-Misner
(ADM) canonical variables for general relativity (GR)
consist of the spatial metric $g_{a b}$ with the conjugate momentum
$p^{ab}$. For bookkeeping purposes we will referred to these variables
as the ``$G$-variables''.
The dynamical evolution is
generated by the action of the form
\begin{equation}\label{SDADM}
\int
\d t
\int
\d^3 x
\left[
{p}^{ab} \dot{g}_{ab} - N \H - N^a \H_a
\right]
\end{equation}
where the over dot denotes a $t$-derivative, $N$ is
the lapse function and $N^a$
the shift vector. This action contains the
the
momentum (diffeomorphism) constraint:
\begin{equation}
\label{Jconstr}
\H_a
=
-2\, \nabla_b \,{p}^b{}_a
\end{equation}
and the Hamiltonian constraint:
\begin{equation}
\label{Hconstr}
\H
=
\mu^{-1} g_{a b c d}\, p^{a b} p^{c d} -  \mu R
\end{equation}
were $\mu := \sqrt{\det g_{ab}}$, $R$ is the Ricci scalar of $g_{ab}$ and
$g_{a b c d} := (g_{a c}g_{b d} + g_{a d}g_{b c} -  g_{a b}g_{c d})/2$.
The Poisson bracket (PB)
with respect to the $G$-variables
$(g_{ab}, p^{ab})$
is denoted by $\PBG{\cdot\,}{\cdot}$.

In a recent paper \cite{Wang2005b} the ADM phase space has been extended to that
consisting of York's mean curvature ``time'' variable $\tau := (4/3) K$, where $K$ is the mean extrinsic curvature,
with
$\mu$ as momentum and conformal metric $\gamma_{a b}$ with
momentum $\pi^{ab}$. Based on York's conformal decomposition of tensors, a
canonical transformation has been found to relate the
$G$-variables to the ``$\Gamma$-variables'' $(\gamma_{a b}, \pi^{a
b}; \tau, \mu)$ via
\begin{equation}\label{ggamma}
g_{a b} = \phi^4 \gamma_{a b} \;\;
\end{equation}
\begin{equation}
\label{pexpr}
p^{a b} = \phi^{-4}\pi^{a b} - \frac12  \, \phi^2 \mu_\gamma\,
\gamma^{ab}\tau
\end{equation}
where $\mu_\gamma := \sqrt{\det \gamma_{ab}}$ and
\begin{equation}\label{phi}
\phi := \left(\frac{\mu}{\mu_\gamma}\right)^{1/6}
\end{equation}
is the conformal
factor. Note that here $\phi$ is regarded as a function of the canonical variables $\mu$ and $\gamma_{ab}$.
It can be seen from the above relations that
a local rescaling of
$\gamma_{ab}$ and $\pi^{a b}$
while holding $\tau$ and $\mu$ leaves ${g}_{a b}, p^{a b}$ invariant.
This redundancy of the $\Gamma$-variables is
offset by the ``conformal constraint'':
\begin{eqnarray}\label{conformconst}
\C = {\gamma}_{ab}\pi^{ab}
\end{eqnarray}
that
generates conformal transformations through its PB
with respect to the $\Gamma$-variables denoted by
$\PBGamma{\cdot\,}{\cdot}$. The tensor $\gamma_{ab}$ will play the role of the conformal metric,
whereas the
original $g_{ab}$ will stay as the physical metric.

The ADM diffeomorphism and Hamiltonian constraints then take the following forms:
\begin{equation}\label{diffconst}
\H_a
=
-2\, \nab_b \,\pi^b{}_a
+\mu\,\tau_{,a}
+4(\ln\phi)_{,a}\, \C
\end{equation}
\begin{equation}\label{HR}
\H
=
\frac1{\mu}\,\pi{}_{ab}\pi^{ab}
-\frac38\, \tau^2 \mu
- \phi^2 \mu_\gamma  R_\gamma + 8\,\mu_\gamma \phi \Delta_\gamma \phi
+\frac\tau2\,\C
-\frac1{2\mu}\,\C^2 .
\end{equation}
The indices of $\pi^{ab}$ and $\pi_{ab}$
are raised or lowed by the conformal metric $\gamma_{ab}$
and its inverse $\gamma^{ab}$ respectively.
Here we have used
the Levi-Civita connection $\nab$,
scalar curvature $R_\gamma$ and
Laplacian $\Delta_\gamma := \gamma^{ab}\nab_a\nab_b$,
associated with the conformal metric $\gamma_{ab}$.

It is useful to introduce the diffeomorphism constraint $\C_a$
for the $\Gamma$-variables by
\begin{equation}\label{diffconst1}
\C_a
:=
\H_a
-
4(\ln\phi)_{,a}\, \C
=
-2\, \nab_b \,\pi^b{}_a+\mu\,\tau_{,a} .
\end{equation}
Using the preservation of the PB relations by the canonical
transformation from the $G$- to $\Gamma$-variables, the
constraints $\H, \C_a$ and $\C$ can be explicitly shown to be
first class \cite{Wang2005b, Wang2005c}.

By using the relation
\begin{equation}\label{pdotg}
p^{ab} \dot{g}_{ab}
=
-\tau \, \dot{\mu}
+
\pi^{ab} \dot{\gamma}_{ab}
+4(\ln\phi)\,\dot{}\, \C
\end{equation}
the canonical action for GR in the $\Gamma$-variables
can be written as
\begin{equation}\label{act_Gamma}
\int
\d t
\int
\d^3 x
\left[
{\pi}^{ab} \dot{\gamma}_{ab}
+  \mu \, \dot{\tau}
- N \H - N^a \C_a - \Lambda \,\C
\right]
\end{equation}
where $\Lambda$ is a Lagrange multiplier
used to effect weakly vanishing of $\C$.
For further background information on the conformal decomposition in GR, see
\cite{Wang2005b, Wang2005c, Wang2005d} and references therein.

\section{Conformal factor as a canonical variable for gravity}

The structure of $\H$ in \eqref{HR} is quite complicated, with
$\phi$ being a highly nonlinear
expression given in \eqref{phi}. If a canonical transformation
can be found that retains $\gamma_{ab}$ as canonical variables while turning $\phi$ into
a canonical variable, then the conformal decomposition for canonical
GR will be greatly simplified. Below we show that this can indeed be
done by explicit construction of new a set of variables satisfying these
criteria. To this end,
consider the quantities $\tilde{\pi}^{ab}$ and $\tilde{\pi}$ given by:
\begin{equation}\label{undpiab}
\tilde{\pi}^{ab} := \pi^{ab}-\frac12\,\phi^6\mu_\gamma\tau\gamma^{ab}
\end{equation}
\begin{equation}\label{undpi}
\tilde{\pi} := -6\,\phi^5\mu_\gamma\tau = -8\,\phi^5\mu_\gamma K
=
4\,\phi^{-1}g_{ab}p^{ab} .
\end{equation}
From \eqref{undpi} we see that
\begin{equation}\label{undtau}
\tau = -\frac16\,\phi^{-5}\mu_\gamma^{-1}\tilde{\pi} .
\end{equation}
Using \eqref{undpiab} and \eqref{undtau} we have
\begin{equation}\label{undpiab1}
\pi^{ab} = \tilde{\pi}^{ab} -\frac1{12}\,\phi\,\tilde{\pi}\,\gamma^{ab} .
\end{equation}
In terms of these quantities, Eqs. \eqref{ggamma} and \eqref{pexpr} become simply
\begin{eqnarray}\label{ggamma1}
g_{ab} &=& \phi^4 \gamma_{ab}
\\
p^{ab} &=& \phi^{-4} \tilde{\pi}^{ab} .
\label{pexpr1}
\end{eqnarray}
It follows that
\begin{equation}\label{}
-\tau\dot\mu
=
\tilde{\pi}\,\dot{\phi}
-\phi^6\tau\dot{\mu}_\gamma
\end{equation}
and
\begin{equation}\label{}
\pi^{ab}\dot{\gamma}_{ab}
=
\tilde{\pi}^{ab}\dot{\gamma}_{ab}
+\phi^6\tau\dot{\mu}_\gamma .
\end{equation}
Therefore the relation
\begin{equation}\label{}
\pi^{ab}\dot{\gamma}_{ab}-\tau\dot\mu
=
\tilde{\pi}^{ab}\dot{\gamma}_{ab}+\tilde{\pi}\,\dot{\phi}
\end{equation}
holds and it shows that the variables
$(\gamma_{ab}, \tilde{\pi}^{ab}; \phi, \tilde{\pi})$
are canonical. We shall refer to them as the ``$\Phi$-variables''
and denote the PB with respect to them by
$\{ \cdot, \cdot\}^\Phi$.
Using \eqref{ggamma1}, \eqref{pexpr1} we see the preservation of the PBs
$\{ \A, \B\}^G = \{ \A, \B\}^\Phi$
for any $\A$ and  $\B$ depending on the
$\Phi$-variables through the $G$-variables.
In particular, the Dirac algebra for $\H$ and $\H_a$ is preserved.

From \eqref{undpiab1} we see that the conformal constraint $\C$
given in \eqref{conformconst} becomes
\begin{equation}\label{conformconst1}
\C
=
{\gamma}_{ab}\tilde{\pi}^{ab} -\frac14\,\phi\,\tilde{\pi} .
\end{equation}
The constraint $\C$
generates the conformal transformation
in the $\Phi$-variables via the following PB relations:
\begin{align}\label{}
\{ \gamma_{ab}(x), \C(x')\}^\Phi &=  \gamma_{ab}(x)\, \delta(x,x') \\
\label{}
\{ \tilde{\pi}^{ab}(x), \C(x')\}^\Phi &= -\tilde{\pi}^{ab}(x)\, \delta(x,x') \\
\label{}
\{\phi(x), \C(x')\}^\Phi &= -\frac14\,\phi(x)\, \delta(x,x') \\
\label{}
\{\tilde{\pi}(x), \C(x')\}^\Phi &= \frac14\,\tilde{\pi}(x)\, \delta(x,x') .
\end{align}
From \eqref{ggamma1} and \eqref{pexpr1} we see that
the physical 3-metric and its momentum
are independent of on $\tilde{\pi}$ and
are conformally invariant:
\begin{align}\label{}
\{g_{ab}(x),  \C(x')\}^\Phi &= 0,\quad \{g_{ab}(x),  \phi(x')\}^\Phi = 0 \\
\label{}
\{p^{ab}(x),  \C(x')\}^\Phi &= 0,\quad \{p^{ab}(x),  \phi(x')\}^\Phi = 0
\end{align}
which, along with the preservation of PBs, imply
\begin{align}\label{}
\{\H(x), \C(x')\}^\Phi &= 0,\quad  \{\H(x), \phi(x')\}^\Phi = 0 \\
\label{}
\{\H_a(x), \C(x')\}^\Phi &= 0,\quad \{\H_a(x), \phi(x')\}^\Phi = 0 .
\end{align}
It follows that $\H, \H_a, \C$ form a set of first class constraints.
To construct the explicit expressions of $\H$ and $\H_a$
in the $\Phi$-variables,
substitute \eqref{ggamma1}, \eqref{pexpr1} and \eqref{conformconst1} into \eqref{Jconstr}
to get
\begin{equation}
\label{Jconstr1}
\H_a
=
-2\, \nab_b \,\tilde{\pi}^b{}_a
+\tilde{\pi}\,\phi_{,a}
+4(\ln\phi)_{,a}\, \C
\end{equation}
and
substitute \eqref{ggamma1}, \eqref{pexpr1} into \eqref{Hconstr}
to get
\begin{equation}
\label{Hconstr1}
\H
=
\phi^{-6}\mu_\gamma^{-1} \gamma_{a b c d}\, \tilde{\pi}^{a b} \tilde{\pi}^{c d}
-  \phi^2 \mu_\gamma R_\gamma + 8\,\mu_\gamma \phi \Delta_\gamma \phi
\end{equation}
where
$\gamma_{a b c d} :=
(\gamma_{a c}\gamma_{b d} + \gamma_{a d}\gamma_{b c} -  \gamma_{a b}\gamma_{c d})/2$.
As with the $\Gamma$-variables, it is useful to introduce the diffeomorphism constraint
in the $\Phi$-variables as
\begin{equation}\label{}
\C_a
=
\H_a
-
4(\ln\phi)_{,a}\, \C
=
-2\, \nab_b \,\tilde{\pi}^b{}_a+\tilde{\pi}\,\phi_{,a}
\end{equation}
and adopt
\begin{equation}\label{act_Phi}
\int
\d t
\int
\d^3 x
\left[
\tilde{\pi}^{ab} \dot{\gamma}_{ab}
+\tilde{\pi}\,\dot{\phi}
- N \H - N^a \C_a - \Lambda \,\C
\right]
\end{equation}
as the canonical action for GR in the $\Phi$-variables.

It follows from the Dirac algebra for the original ADM constraints
that the complete set of independent constraints $\{\C,\C_a,\H\}$
is indeed first class, satisfying the following algebra:
\begin{align}
\label{}
\{\C(x),  \C(x')\}^\Phi
&=
0
\\
\label{}
\{\C_a(x), \C_b(x') \}^\Phi
&=
\C_b(x)\,\delta_{,a}(x, x')
-
(a x\leftrightarrow b x')
\\
\label{}
\{\C_a(x),  \C_(x')\}^\Phi
&=
\C(x)\, \delta_{,a}(x, x')
\\
\label{}
\{\H(x), \H(x') \}^\Phi
&=
\phi^{-4}\gamma^{ab} (\C_a + 4(\ln\phi)_{,a}\, \C)(x)\,\delta_{,b}(x, x')
-
(x\leftrightarrow x')
\\
\label{}
\{ \C(x),  \H(x')\}^\Phi
&=
0
\\
\label{}
\{\C_a(x), \H(x') \}^\Phi
&=
 \H(x)\, \delta_{,a}(x, x') .
\end{align}
The above new formulation can be applied to
simplify the  spin-gauge treatment in
\cite{Wang2005b, Wang2005c}
and to develop the conformal
approach to loop quantum gravity.

\begin{acknowledgments}
I would like to thank
R. Bingham,
S. Carlip,
J. T. Mendon\c{c}a,
N. O'Murchadha
and
J. W. York for stimulating discussions and
the Aberdeen Centre for Applied Dynamics Research and the
CCLRC Centre for Fundamental Physics for financial support.
\end{acknowledgments}


\end{document}